\documentclass[prd,aps,preprint,epsf,epsfig,floats,nofootinbib]{revtex4}
\usepackage{graphics}
\usepackage{epsfig}

\usepackage{graphicx}

\usepackage[unicode=true,pdfusetitle,
 bookmarks=true,bookmarksnumbered=false,bookmarksopen=false,
 breaklinks=true,pdfborder={0 0 0},backref=false,colorlinks=true]
 {hyperref}

 \def\lsim{\mathrel{\vcenter{\hbox{$<$}\nointerlineskip\hbox{$\sim$}}}}
\def\gsim{\mathrel{\vcenter{\hbox{$>$}\nointerlineskip\hbox{$\sim$}}}}
 
 \newcommand{\llnunu}{ \ell \bar{\ell} \nu \bar{\nu }}
  \newcommand{\gev}{ \textrm{ GeV} }

\begin{document}

\preprint{\vbox{ \hbox{UMD-PP-012-012} }}
\title{\Large\bf Bounds on TeV Seesaw Models from LHC Higgs Data }
\author{\bf P. S. Bhupal Dev$^{1,2}$, Roberto Franceschini$^1$ and R. N. Mohapatra$^1$}
 \affiliation{$^1$Maryland Center for Fundamental
Physics and Department of Physics, University of Maryland, College
Park, MD 20742, USA.\\
$^2$ Consortium for Fundamental Physics, School of Physics and Astronomy, 
University of Manchester, Manchester M13 9PL, United Kingdom.}

\begin{abstract}
We derive bounds on the Dirac Yukawa couplings of the neutrinos in seesaw models using the recent Large Hadron Collider (LHC) data on 
Higgs decays for the case 
where the Standard Model (SM) singlet heavy leptons needed for the seesaw mechanism have masses in the 100 GeV range. Such scenarios with 
large Yukawa couplings are natural in Inverse Seesaw (ISS) models since the small neutrino mass owes its origin to a small Majorana mass 
of a new set of singlet fermions. Large Yukawas with sub-TeV mass right-handed neutrinos are also possible for certain textures in Type-I 
seesaw models, so that the above bounds also apply to them. We find that the current Higgs data from the LHC can put bounds on both 
electron- and muon-type Yukawa couplings of order $10^{-2}$.
  \end{abstract}
\maketitle

\section{Introduction}
Understanding the new physics behind small neutrino masses is a major current focus of particle theory research. The simplest scenarios for this are the various types of seesaw mechanisms, which have also formed the basis for understanding  the observed mixing pattern, which is so different from the quark sector. Two of them, type-I~\cite{seesaw} and the inverse seesaw~\cite{inverse} postulate the existence of heavy SM singlet fermions, generally denoted by the symbol $N_a$ ($a$ is a generation index) in addition to the SM fermions. The stability of these patterns can be guaranteed by extending the SM gauge group as we discuss below. These singlet fermions couple to familiar lepton doublets via Yukawa couplings of the form ${\cal L}_Y=y_{ab}\bar{L}_a H N_b+{\rm h.c.} $, which after electroweak symmetry breaking leads to the Dirac mass of the neutrino, which is expected to be in the GeV range. 
In the type-I seesaw, the tiny neutrino mass arises once we introduce an additional Majorana mass term $M_{ab}N_aN_b$ for the $N$'s. Being SM singlets, these Majorana particles can take arbitrarily 
large mass values, which then guarantees that the neutrino mass given by the formula below is small: 
\begin{eqnarray}
M_\nu=-yM^{-1}y^T v^2_{\rm wk}\,.
\end{eqnarray}
Note that since the scale of neutrino masses is known to be in the sub-eV range, the scale $M$ is correlated with the magnitude of the Yukawa couplings $y_{ab}$. For instance, if the $N$ masses are in the 100 GeV range, it implies $y\sim 10^{-5}$ or so. However, for specific flavor pattern of both $M_{ab}$ and  $y_{ab}$, tiny neutrino masses can be realized without making $y$ necessarily tiny. Examples of such theories are given in~\cite{KS,others}. 

On the other hand, in the case of inverse seesaw models, in addition to the set of singlets $N_a$, one adds another set of singlet fermions $S_a$ which form a Dirac mass $M$ with the $N_a$ fields, i.e., $M_{ab}N_aS_b$. The model so far has conserved lepton number and leads to zero neutrino mass~\cite{WW}. One then allows the $S$ fields to have a Majorana mass matrix $\mu_S$ whose overall scale is in the keV range~\cite{inverse}. 
The neutrino mass formula in this case is given by:
\begin{eqnarray}
	M_\nu=y({M^T})^{-1}\mu_S M^{-1}y^T v^2_{\rm wk}  \label{massISS} \,.
\end{eqnarray}
This also leads to tiny Majorana mass for neutrinos. The small scale of the $\mu_S$ matrix can be explained in various ways~\cite{fong} and is a window to further new physics. In this model since the small neutrino masses arise from the small values of $\mu_S$ matrix elements, the singlet neutrinos $(N, S)$ form a Dirac pair with masses in the 100 GeV range and yet have Yukawa couplings of order ${\cal O}(1)$ without any fine-tuning.

A key question for neutrino mass physics is how to test the seesaw mechanism~\cite{han}. The obvious first step would be to determine the couplings $y_{ab}$ and $M_a$. The vast literature on neutrino mass models is devoted to precisely this question where additional theoretical assumptions such as symmetries are used for this purpose. The symmetries restrict the theory to a particular sub-space of the full parameter-space. If the seesaw scale is high, as is the case in most type-I models, this appears to be the only realistic possibility. We take a different approach in this paper. Since $y_{ab}$ denotes the coupling of the $N$-fermions to the Higgs boson, if $M$ is in the 100 GeV range, experimental information on $y$ can be obtained from the LHC data on the Higgs boson decays~\cite{carp}. Our goal in this paper is to focus on this question. 

In this paper, we  consider primarily  the inverse seesaw case and   we will comment on the specific type-I scenarios where our results will apply with slight changes. The main result of the paper i.e. bounds on the the Yukawa couplings $y_{ab}$, is derived from an analysis of the 7-TeV LHC data. The 8-TeV data came out after this work was finished. However, we have redone our analysis with this new data and found that the change in the bounds is very minimal. We comment on it briefly at the end of the paper. 

This paper is organized as follows: in Section \ref{gaugemodel}, we discuss the minimal gauge models with inverse seesaw; in Section \ref{generalpheno} we give an overview of the Higgs phenomenology of the seesaw models; in Section \ref{llnunubound} we shall illustrate how to constrain the parameter-space in 
seesaw models using the results of the Higgs searches in the final state $\ell\bar{\ell}\nu\bar{\nu}$; in Section \ref{postdiscovery}  we derive these 
constraints on the seesaw models by analyzing the rates of all the measured decay modes of the Higgs boson; the concluding remarks are in Section \ref{conclusion}. In Appendix A, we list the relevant decay widths of the heavy neutrino.  

\section{Gauge models with large Yukawa couplings and sub-TeV singlet fermions \label{gaugemodel}}
In this section, we briefly discuss two classes of model which can have ${\cal O}(1)$ Dirac Yukawa couplings and $\sim 100$ GeV mass for the singlet fermions 
responsible for the seesaw mechanisms\footnote{There are other classes of models, e.g., linear seesaw~\cite{linear} and double 
seesaw~\cite{inverse,ma09} models which can also have this feature.}. One class, as noted above, is the inverse seesaw model, where this situation can be naturally realized, and a second class, which belongs to  type-I seesaw-based neutrino mass models where with specific texture for the Dirac Yukawa matrices as well as for the singlet fermion mass matrix, one can have LHC-accessible parameters.  

\subsection{Inverse seesaw gauge model} 
If we simply add two heavy SM singlet fermions $N_a, S_a$ to the SM, there are many gauge invariant terms in the potential and we will not get the simple inverse seesaw formula. The simplest extension of SM where the inverse seesaw formula arises  naturally is $SU(2)_L\times U(1)_{I_{3R}}\times U(1)_{B-L}$ with gauge symmetry under which the singlet fermions $N$  transform as $(1,+1/2, -1/2)$ and $S$ fields are singlets. The gauge symmetry is broken by a Higgs field $\chi (1, +1/2, -1/2 )$ acquiring vacuum expectation value (vev) and the SM-doublet $H$ vev then breaks the gauge 
group down to electromagnetic $U(1)$. The Yukawa Lagrangian for the lepton sector is given by:
\begin{eqnarray}
	{\cal L}_Y= y_{ab}\bar{L}_aHN_b+f_{ab} \bar{N}_a\chi S_b +\mu_{S_{ab}}S_aS_b +{\rm h.c.} \label{lagrangianISS}
\end{eqnarray}
It is clear that once we substitute $\langle H^0\rangle =v_{\rm wk}\equiv v/\sqrt{2}=174 \textrm{ GeV}$ and $\langle \chi \rangle = v_{BL}$, we get the inverse seesaw matrix~\cite{inverse}
\begin{eqnarray}
M_{\nu,N,S}~=~\left( \begin{array}{ccc}0 & yv/\sqrt{2} & 0\\ y^Tv/\sqrt{2} & 0 & fv_{BL}\\ 0 & f^Tv_{BL} & \mu_S\end{array}\right)
\end{eqnarray}
with the mass matrix for the neutrinos given by Eq. (2) with $M=fv_{BL}$ (Note that $y,f,\mu_S$ are in general $3\times 3$ matrices).
At the renormalizable level, there are no other terms allowed in the Yukawa Lagrangian by the gauge symmetry and therefore the above inverse seesaw formula for neutrinos is stable under radiative corrections. 

Our goal is to find constraints on $y$ as a function of $M$. There exists an extensive analysis on the constraints for light singlet fermions~\cite{atre}. Although this analysis addresses only the constraints on Majorana neutrinos, some of their results apply to our case as well. However, the limits derived in~\cite{atre} for $M> 60 $ GeV or so are very weak. Furthermore constraints from neutrino-less double beta decay~\cite{beta} derived  on heavy sterile neutrinos do not apply to this case since in our model, the $N$ and $S$ form a pseudo-Dirac pair and lepton number is almost exactly conserved.

In order to use the LHC data to explore constraints on $y$ and $M$ in the 100 GeV range, we will assume that (i) $v_{BL} \gg v_{\rm wk}$ and (ii)  the mass of Re($\chi^0$) is heavy compared to the SM Higgs boson so that neither the heavy gauge boson associated with $(B-L)$-symmetry nor the interactions of 
Re($\chi^0$) affect the Higgs boson decay modes we consider.

\begin{figure}
	\centering
	\includegraphics[width=8cm]{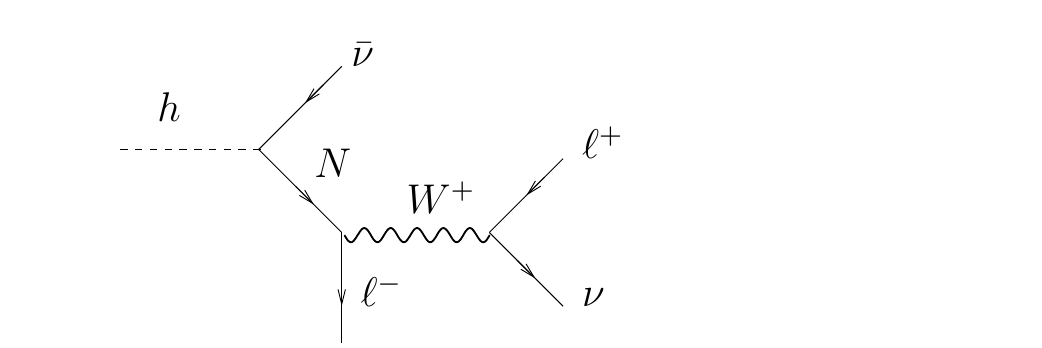}
	\includegraphics[width=8cm]{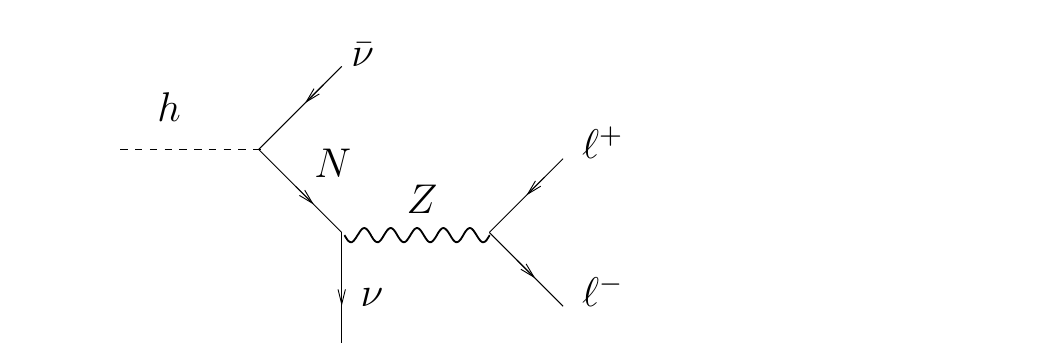}
	\caption{The Higgs decay modes into $2\ell 2\nu$  mediated by the ISS couplings.}
	\label{fig:hdecayISS}
\end{figure}

It follows from the above Lagrangian that if one of the singlet fermions has mass in the 100 GeV range, it will affect the Higgs branching ratios: for instance if $M_N < M_h$, then this opens up a new mode for SM Higgs decay, i.e., $h\to \bar\nu_a N_b$, and the collider signal will arise from $N-\nu$ mixing diagram 
in Fig.~\ref{fig:hdecayISS} where $N\to \nu Z, \ell W$.
Folding $W, Z$ decays, one will get final states with $\nu\bar\nu\ell_a\ell_b$ where in the final state both charged leptons and anti-leptons will appear 
and the existing LHC data on these final states will provide constraints on $y$. Clearly, which charged lepton appears will depend on the flavor structure of $y$ and $f$. For $f$ we will go to a basis so that it is diagonal, i.e. a linear combination of $\nu$ and $N$  are mass eigenstates with $S$ field providing the chiral Dirac partner.

\subsection{Type-I seesaw case}
Turning to the type-I case, as noted earlier, in generic models, the Dirac Yukawa couplings are very small for the seesaw scale in the TeV regime. However, for specific textures for $y$, it is possible to attain singlet fermion mass in the 100 GeV range with Dirac Yukawa $y$'s of order ${\cal O}(1)$ while still
satisfying the neutrino oscillation data. In this case the singlet fermions could show up at the LHC. Two examples of this type of texture are~\cite{KS}:
\begin{eqnarray}
y_{ab}~=~y^0_{ab} + \delta y_{ab} ~~~~~\textrm{with} \\
y^0_{ab}~=~\left(\begin{array}{c}1 \\ \alpha \\ \beta\end{array}\right)\left(\begin{array}{ccc} y_1 & y_2 & y_3\end{array}\right)
\end{eqnarray}
with the constraint that $\frac{y^2_1}{M_1}+\frac{y^2_2}{M_2}+\frac{y^2_3}{M_3}=0$ where $M_a$ are the singlet neutrino masses and the singlet fermion mass matrix is diagonal. Seesaw of this kind leads in the leading order to zero neutrino masses; small $\delta y_{ab}$ can then generate the neutrino masses and mixings, and this will not affect the Yukawa couplings or the masses $M_a$.

A second example has $y_{ab}$ of the following type~\cite{KS}:
\begin{eqnarray}
y_{ab}~=~\left(\begin{array}{ccc} y_a & \delta_a & \epsilon_a \\y_b & \delta_b & \epsilon_b \\y_c & \delta_c & \epsilon_c\end{array}\right) \, ,
	\label{eq:democratic}
\end{eqnarray}
with right handed neutrino mass matrix of the form
\begin{eqnarray}
M~=~\left(\begin{array}{ccc} 0&M_1 & 0 \\ M_1 & 0 & 0 \\ 0 & 0 & M_2\end{array}\right) \, ,
\end{eqnarray}
with $\epsilon_{a,b,c} , \delta_{a,b,c} \ll y_{a,b,c} $. In the limit of $\epsilon, \delta =0$ the neutrino mass matrix is a null matrix regardless of the values of $M_a$  and $y_{a,b,c}$ and small neutrino masses arise from choosing $\epsilon$'s and $\delta$'s small. To the best of our knowledge, we are not aware of any symmetries that will guarantee these textures for the $M$ and $y$ matrices and it will be interesting to seek symmetry origin for them .
Therefore again in this case, LHC data should put constraints on $y_{a,b,c}$ vs $M_a$. Since in this case, $y$'s are arbitrary, they could be of democratic type which will mean that there are multiple flavor final states.

Two important points regarding these models are worth emphasizing. In both models, as the small parameters ($\mu_S$ in the case of inverse seesaw, and $\delta$ and $\epsilon$ in the type-I model discussed above) go to zero, the neutrinos become massless. As a result, the neutral Higgs decays to two modes: $h\to \nu+\bar{N}, \bar{\nu}+N$. Unlike the case where both $\nu$ and $N$ are Majorana fermions (as in generic type-I seesaw models), the Higgs decay rate to $\nu+N$ final states becomes twice as large. Thus, the discussions of Higgs decay in both classes of models are similar except for the flavor richness in case (B) compared to case (A) . 
In case (B) model (II), for $y_a\sim y_b \sim y_c$, all flavors couple with equal strength to a single fermion flavor $N_a$ implying that final state signal  becomes a combination of all three flavors. Similar situation also occurs  for case (B), model I if $\alpha, \beta \sim 1$. 

\section{Seesaw Higgs phenomenology\label{generalpheno}}

As discussed above, a new Yukawa interaction in the leptonic sector can potentially alter
the Higgs phenomenology in a dramatic way. In fact, in the models considered
here, the Yukawa coupling of the neutrino can be sizable -- in principle,
much larger than the largest Yukawa coupling involved in the decay of a light SM Higgs boson, i.e., the
one with the bottom quark which is of order ${\cal O}(10^{-2})$.

Here we shall describe the observable consequences of the new Yukawa coupling. 
Focusing on the inverse seesaw case, it is apparent from the neutrino mass formula in Eq.~(\ref{massISS}) that neutrino mass fits will largely be dictated by the combination of the flavor structure of the $\mu_S$ matrix and the $y$-matrix. 

In quark-lepton unified theories, the natural expectation is that flavor mixings in $y$ are ``weak''. So we can assume $y$ to be a diagonal matrix. The next question is the relative magnitude of different flavor Yukawa couplings. Below we will consider the cases when (a) $y_{\nu_{e}} \gg y_{\nu_{\mu}} \gg y_{\nu_{\tau}}$, (b) $y_{\nu_{\mu}} \gg y_{\nu_{e}} \gg y_{\nu_{\tau}}$, and (c) a democratic form where all the Yukawa couplings are equal. These different coupling structures imply distinctive flavor structure of the final states. 
From the interaction Lagrangian, Eq.~(\ref{lagrangianISS}), 
it follows that $y_{ab}$ mediates the decays illustrated in Fig.~\ref{fig:hdecayISS}:
\begin{equation}
	h\to\bar{\nu}_{a}N_{b}+{\rm c.c.}\to\bar{\nu}_{a}\nu_{b}Z+\left(\bar{\nu}_{a}\ell_{c}^{-}W^{+}+{\rm c.c.}\right)\,,\label{eq:GenericDecays}
\end{equation}
where charged and neutral leptons arise in all the flavor combinations allowed by the form of the $y$ matrix.
The gauge bosons $Z$ and $W$ arise in the decay of $N$ through the $\nu-N$ mixing, that is ${1 \over \sqrt{2} }yv/m_{N_{b}}$. The gauge bosons in turn can decay into  leptons or
hadrons, such that the new Yukawa coupling contributes to the rates of both 
fully leptonic and semi-leptonic final states. 

The heavy particles $N_{b}$,
$W$ and $Z$ in the decay chain of Eq.~(\ref{eq:GenericDecays}) can be either on-shell
or off-shell depending on the mass of $N_{b}$. In particular, when $m_{h}>m_{N_{b}}$ the Higgs decay width in Eq.~(\ref{eq:GenericDecays}) scales as $y^{2}$ while for $m_{h}<m_{N_{b}}$, it scales as $y^{4}$.

The fully leptonic decay mode of the Higgs $\ell\bar{\ell}\nu\bar{\nu}$
is among the most sensitive search channels of the Higgs and therefore
is a very sensitive probe of the $y$-coupling, especially
for $m_{N_b}$ not far above or below $m_{h}$. This allows us to derive a bound on the $(m_{N_b},y)$ parameter space as described in detail in section~\ref{llnunubound}.

For definiteness in what follows, we shall  focus on the  case of 
the ISS  model with only one light flavor
of heavy neutrino which we choose to be the electron flavor, $N_{e}$, 
and the corresponding Yukawa coupling $y_{\nu_{e}}$ to the Higgs doublet. 
We generically denote the mass of the heavy neutrino as $m_{N}$, suppressing 
the explicit flavor index $e$.

In this scenario the decay chain of Eq.~(\ref{eq:GenericDecays}) becomes
\begin{equation}
	h\to\bar{\nu}_{e}N_{e}+{\rm c.c.}\to\bar{\nu}_{e}\nu_{e}Z+\left(\bar{\nu}_{e}e^{-}W^{+}+{\rm c.c.}\right)\,.\label{eq:ISSdecays}
\end{equation}

The decay of the $Z$ boson to leptons produces pairs of charged leptons of all flavors, and hence, it contributes to the same flavor final states, namely 
$$h\to e^{-}e^{+}\bar{\nu}_{e}\nu_{e}\textrm{ and }h\to \mu^{-}\mu^{+}\bar{\nu}_{e}\nu_{e}\,.$$

The decay of the $W$ boson produces a single charged lepton. In this case, the $W$ is produced by a mixing effect of the heavy $N_{e}$, and therefore, the $W$ is always produced in 
association with an electron or a positron. Hence, the decay process mediated 
by the $W$ only contributes to electron final states, namely 
$$ h \to e^{-}\mu^{+}  \bar{\nu}_{e} \nu_{\mu} + {\rm c.c.}\textrm{ and  }h \to e^{-} e^{+} \bar{\nu}_{e} \nu_{e}\,. $$

Altogether, due to the assumed coupling  structure $y_{\nu_{e}} \gg y_{\nu_{\mu}} \gg y_{\nu_{\tau}}$, the final state leptons of the Higgs decays mediated by the ISS are both of opposite flavor and of same flavor, with same flavor $\mu$ 
getting its only contribution from processes with $Z$-boson intermediate state.


For the other coupling patterns the resulting flavor structures are straightforward modifications of the one considered here. For the pattern (b) where $y_{\nu_{\mu}} \gg y_{\nu_{e}} \gg y_{\nu_{\tau}}$, the main difference is that now it is the same flavor $e$ final state that gets its only contribution from the $Z$-intermediate state. The opposite flavor  final states again get contribution from both $Z$- and $W$-mediated processes, though with the kinematics  of $e$ and $\mu$ exchanged. 

As in our analysis described in the next sections there is going to be little difference between $e$ and $\mu$, the bounds derived for the ISS with light $N_{e}$ also apply for the case of light $N_{\mu}$. 

So far we considered a basis in which the $f$ matrix is diagonal. There are however possible situations where one can derive bounds on Yukawa couplings. For instance, it is possible to have structure of the $f$ and $y$ matrices where democratic Yukawa couplings can emerge i.e. by assuming an appropriate  texture for the $NS$  Dirac mass matrix we can have the lightest heavy neutrino 
$N$ as a linear combination of all the three flavor states with 
equal probability, i.e., $N_1=(N_e+N_\mu+N_\tau)/\sqrt 3$. In this case, if  $y$ is a unit matrix, then we expect all the possible flavor combinations to be equally populated as both the interactions of $N_{a}$ and the gauge bosons are flavor-universal. The analyses of the LHC experiments are mostly sensitive to $e$ and $\mu$ flavors only, therefore for such a  democratic case, only 4 out the 9 possible final state flavor combinations are easily detectable~\footnote{In principle, the leptonic decays of the $\tau$ may be used to recover some sensitivity. However, they are only a fraction of the $\tau$ decay modes, and there are detection inefficiencies. As such, the gain might not be huge. At any rate, the use of the $\tau$'s requires careful experimental considerations that go beyond our competence; hence we prefer not to consider them.}. Furthermore the effect of the mixing of flavors in the lightest mass eigenstate $N$ in general reduces the rate of $e$ and $\mu$ leptons. Because of these effects the bounds should be recalculated carefully to account for the different flavor structure of the signal. 

However, neglecting the sub-dominant processes mediated by the $Z$, one can obtain an estimate for the bound in the democratic case. Indeed, the rates of the processes mediated by the $W$ are just rescaled by mixing and multiplicity factors compared to those of the ISS model with only a light $N_{e}$. As such, the bounds of the ISS case should be relaxed by a constant factor. In particular, from the fact that the rate of  Eq.~(\ref{eq:ISSdecays}) scales as $y_{\nu_{e}}^{2}$ when $m_{h}>m_{N}$ and as $y_{\nu_{e}}^{4}$ when $m_{h}<m_{N}$, we find that the bound gets relaxed by a factor $(3/2)^{1/2}$ or $(3/2)^{1/4}$ in the two cases. 

In the following, we shall illustrate in detail how to derive a bound on the Yukawa coupling in the ISS model. Using the same procedure, we also computed the bound for the type-I democratic case and we checked that this simple  rescaling argument describes the actual bound to a very good accuracy. Hence, we give only the bounds for the ISS case, and those for the democratic type-I case can be obtained by the simple rescaling described above.

%

Of course a more targeted analysis from the LHC experiment with flavor structure explicitly taken into account could distinguish between these coupling structures. For instance, an imbalance between the $e$ and $\mu$ same flavor final states can distinguish case (a) and case (b), while case (c) would be favored if all the flavor combination appear to be equally populated.

Irrespective of the flavor structure of the Yukawa interactions, the new Yukawa coupling mediates new decay modes of the Higgs, which contribute to its total width, and hence, in these seesaw models, the Higgs width is larger than in the SM:
\[
\Gamma_{h}=\Gamma_{\rm SM}+\Gamma_ {\rm seesaw}\,.
\]
The seesaw contribution to the the total width is necessarily model-dependent. For the case of the ISS with dominant coupling $y_{\nu_{e}}$, the width \footnote{This formula sums $h \to \nu \bar{N}$ and $h \to \bar{\nu} N$.} is given by 
\begin{equation}
\Gamma_{\rm ISS}=\frac{y_{\nu_{e}}^2 }{8 \pi  m_{h}^3}\left(m_{h}^2-m_{N}^2\right)^2\, , \label{GammaISS}
\end{equation}
where we have normalized the Yukawa couplings such that the Dirac mass terms of all fermions are $m_{D}={y v \over  \sqrt{2}}$ for $v\simeq 246 \gev$.
In the case of the democratic type-I seesaw, as in Eq.~(\ref{eq:democratic}), we can take $y_{\rm demo}\equiv y_{a}=y_{b}=y_{c}$; hence the Lagrangian 
essentially contains three interactions of equal strength for the Higgs decay, and the Higgs decay width is the same as in the ISS model once 
$y_{\rm demo}=y_{\rm ISS}/3$ is taken.

In Section \ref{postdiscovery} we shall discuss in detail how to use the information on the measured rates of the several Higgs decay final states to put a bound on the seesaw coupling.


\section{Bounds from the search of the Higgs boson in the $\ell\bar{\ell}\nu\bar{\nu}$
final state  \label{llnunubound}}

The ATLAS and CMS collaboration have both found evidence for a Higgs-like particle at around 125 GeV. The main evidence for the new particle comes from final states with  resonant two photons or  four leptons \cite{HiggsJuly2012,CMS-PAS-HIG-12-020,ATLAS-CONF-2012-093}. 
In addition to the $4\ell$ and $\gamma\gamma$ searches the LHC and TeVatron experiments searched for a SM Higgs
boson in several other final states, including in the final state  $\llnunu$ \cite{CMS-Collaboration:2012vn, CMS-PAS-HIG-12-017,ATLAS-Collaboration:2012kx,The-CMS-Collaboration:2012pd,ATLAS-Collaboration:2012mz}.

The experiments presented cut-based as well as multivariate analyses
to put a bound on the mass of the SM Higgs boson. The bounds from
the multivariate analysis are generically (slightly) more stringent
than the one obtained from the cut-based analysis. However the multivariate
analysis cannot be easily reproduced with our means, and hence, we shall only use the cut-based analysis to derive our bound.

Here we shall reinterpret the results of~\cite{CMS-Collaboration:2012vn} to extract
a bound on extra sources of $\llnunu$ events. To do this, we shall repeat the 
cut-based analysis of~\cite{CMS-Collaboration:2012vn}  on event samples generated by Monte Carlo tools -- 
matrix elements computed with $\textsc{Madgraph5}$~\cite{Alwall:2011fk}, 
showered and hadronized with $\textsc{PYTHIA6.4}$~\cite{Sjostrand:2006zr}
and detector response parametrized by $\textsc{Delphes1.9}$~\cite{Ovyn:2009ys}.
Hadrons have been clustered into jet with the anti-kT algorithm as
implemented in $\textsc{FastJet2}$~\cite{Cacciari:2011rt,Cacciari:2006vn}.

%
%
%
%


In~\cite{CMS-Collaboration:2012vn}, the CMS collaboration performed several analysis on the $\llnunu$ sample collected in the year
2011 with the LHC running at $7$ TeV center-of-mass energy. In particular
they made a basic selection on the leptons, jets and missing energy
of the events depending on the flavor of the final state leptons.
The cuts for the opposite flavor (OF) and same flavor (SF) cases are
reported in Table \ref{tab:Baseline-selection}. Then the analysis
is specialized for specific values of the SM Higgs boson mass and
further cuts are devised. These cuts are collected in the Tables \ref{tab:A120}
and \ref{tab:A130} for the Higgs mass hypothesis of $120$ GeV and $130$
GeV, respectively. These analyses, which we call $A_{120}$ and $A_{130}$,
are the most sensitive to new physics connected to the Higgs-like particle recently discovered, and hence,  
these are the only analyses we are going to repeat in order to extract our bound. 

\begin{table}
\begin{centering}
\begin{tabular}{|c|c|c|}
\hline 
\multicolumn{3}{|c|}{Baseline selection for all $m_{h}$}\tabularnewline
\hline 
$OF(e\mu)$  & $SF(\mu\mu)$ & $SF(ee)$\tabularnewline
\hline 
\hline 
$n_{\mu}=1$, $n_{e}=1$ & $n_{\mu}=2$ & $n_{e}=2$\tabularnewline
\hline 
$|\eta_{e}|<2.5$, $|\eta_{\mu}|<2.4$ & $|\eta_{\mu}|<2.4$ & $\eta_{e}<2.5$\tabularnewline
\hline 
$\Delta R_{\ell\ell}>0.3$ & $\Delta R_{\mu\mu}>0.3$ & $\Delta R_{ee}>0.4$\tabularnewline
\hline 
mET $>$ 20 GeV & \multicolumn{2}{c|}{mET$>$40 GeV}\tabularnewline
\hline 
$m_{\ell\ell}>12$ GeV & \multicolumn{2}{c|}{$m_{\ell\ell}>20$ GeV}\tabularnewline
\hline 
\multicolumn{3}{|c|}{$p_{T,\ell\ell}>45$ GeV}\tabularnewline
\hline 
\end{tabular}
\par\end{centering}

\caption{\label{tab:Baseline-selection}Baseline selection of the CMS cut-based
analysis, as in~\cite{CMS-Collaboration:2012vn}. }

\end{table}
\begin{table}
\begin{centering}
\begin{tabular}{|c|c|c|}
\hline 
\multicolumn{2}{|c|}{Analysis $A_{120}$ (tailored for SM Higgs $m_{h}=120$ GeV)}\tabularnewline
\hline 
$OF(e\mu)$  & $SF(ee,\mu\mu)$\tabularnewline
\hline 
\hline 
$p_{T,\ell_{2}}>10$ GeV & $p_{T,\ell_{2}}>15$ GeV\tabularnewline
\hline 
\multicolumn{2}{|c|}{$p_{T,\ell_{1}}>20$ GeV }\tabularnewline
\hline 
\multicolumn{2}{|c|}{$m_{\ell\ell}<40$ GeV}\tabularnewline
\hline 
\multicolumn{2}{|c|}{$\Delta\phi<115^\circ$}\tabularnewline
\hline 
\multicolumn{2}{|c|}{$m_{T,\ell\ell mET}\in [80,120]$ GeV}\tabularnewline
\hline 
\end{tabular}$\quad$
\par\end{centering}

\caption{\label{tab:A120}Cuts added to those in Table \ref{tab:Baseline-selection}
in the analysis devised in~\cite{CMS-Collaboration:2012vn} for the hypothesis $m_{h}=120$ GeV.}
\end{table}
\begin{table}
\begin{centering}
\begin{tabular}{|c|c|c|}
\hline 
\multicolumn{2}{|c|}{Analysis $A_{130}$ (tailored for SM Higgs $m_{h}=130$ GeV) }\tabularnewline
\hline 
$OF(e\mu)$  & $SF(ee,\mu\mu)$ \tabularnewline
\hline 
\hline 
$p_{T,\ell_{2}}>10$ GeV & $p_{T,\ell_{2}}>15$ GeV\tabularnewline
\hline 
\multicolumn{2}{|c|}{$p_{T,\ell_{1}}>25$ GeV }\tabularnewline
\hline 
\multicolumn{2}{|c|}{$m_{\ell\ell}<45$ GeV}\tabularnewline
\hline 
\multicolumn{2}{|c|}{$\Delta\phi<90^\circ$}\tabularnewline
\hline 
\multicolumn{2}{|c|}{$m_{T,\ell\ell mET}\in [80,125]$ GeV}\tabularnewline
\hline 
\end{tabular}
\par\end{centering}

\caption{\label{tab:A130}Cuts added to those in Table \ref{tab:Baseline-selection}
in the analysis devised in~\cite{CMS-Collaboration:2012vn} for the hypothesis $m_{h}=130$ GeV.}

\end{table}


To obtain a bound on the ISS model we impose that the events yield
of the ISS after the cuts of both the analyses $A_{120}$ and $A_{130}$
is not larger than the total number of events allowed by each of the analysis.
This is done as follows: for each hypothesis for $m_{h}$ in the ISS
model we compute the yield of events after the cuts of CMS. This depends
on the mass $m_{N}$ and coupling $y_{\nu_{e}}$ and it is given by
\begin{eqnarray}
n_{\rm ISS}(m_{N},\, y_{\nu_{e}}) & =L\cdot\sigma_{h} & \left[\epsilon_{\rm SM}\frac{\Gamma(h\to WW^{*}\to\llnunu)}{\Gamma_{\rm SM}+\Gamma_{\rm ISS}}
+\sum_{j,k}\epsilon_{jk}\frac{\Gamma(h\to\bar{\nu}N_{e}+{\rm c.c.}\to\bar{\nu}_{e}\ell_{j}\bar{\ell}_{k}\nu)}{\Gamma_{\rm SM}+\Gamma_{\rm ISS}}\right] \nonumber\\
\label{eq:nISS}
\end{eqnarray}
where $L=4.6~{\rm fb}^{-1}$ is the integrated luminosity used in the analysis,
$\sigma_{h}$ is the total Higgs production cross-section taken from 
~\cite{LHC-Higgs-Cross-Section-Working-Group:2011fj}, $j$ and $k$
are flavor indexes $e,\mu$, and $\epsilon_{\rm SM}$ and $\epsilon_{jk}$
are the efficiencies of the CMS selections for the decays mediated
by the SM decay channel $WW^{*}$ and by decays of the ISS, respectively. 

As we just want to illustrate here how to obtain an upper bound on the Yukawa, 
we shall use only a few representative values of $m_{N}$, namely 60, 100,
140 and 200 GeV. For lighter $m_{N}$ the LHC searches tends to be
rather ineffective. In fact, in the ISS signal both the charged leptons
originate from $N\to\ell_{j}\bar{\ell}_{k}\nu$ and therefore
the invariant mass of the two leptons cannot exceed $m_{N}$. The
cuts $m_{\ell\ell}>12\,(20)$ GeV for OF (SF) leptons, needed to reject
leptons from QCD decays, would remove completely the ISS contribution
for $m_{N}<12\,(20)$ GeV. Also, for light $m_{N}$ the bounds from other
experiments are more stringent~\cite{atre}. The power of the LHC is the
sensitivity to $m_{N}$ around and above the mass of the Higgs, 
which improves significantly as compared to the reach of 
previous direct bounds. 

We remark that our computation of the total events yield neglects the
possible interference between the ISS and the SM contributions, which
in general is a small effect due to different flavor and Lorentz structure
of the decays. 

The selection efficiencies $\epsilon_{\rm SM}$ and $\epsilon_{jk}$ have
been computed with showered events passed through $\textsc{Delphes1.9}$
and hadrons clustered into jets using the anti-kT algorithm with jet-cone 
radius parameter
$R=0.5\,$, as done in the CMS analysis. The obtained efficiencies
have been rescaled such as to reproduce the SM Higgs boson event yield
in Table 3 of~\cite{CMS-Collaboration:2012vn} for $m_{h}=120$ and
$130$ GeV for the analysis $A_{120}$ and $A_{130}$ respectively.
The results of the used chain of simulation codes are rather realistic,
indeed the rescaling factor is almost flat w.r.t the cuts and the
differences between the CMS and $\textsc{Delphes1.9}$ efficiencies are within 20\%.

For the total width of the SM Higgs boson $\Gamma_{\rm SM}$ and the partial
width $\Gamma(h\to WW^{*}\to\llnunu)$ we take the reference values
of~\cite{LHC-Higgs-Cross-Section-Working-Group:2011fj}. 

The width $\Gamma_{\rm ISS}$ due to the decays mediated by $y_{\nu_{e}}$
and the partial widths $\Gamma(h\to\bar{\nu}N_{e}+{\rm c.c.}\to\bar{\nu}_{e}\ell_{j}\bar{\ell}_{k}\nu)$
have been computed with $\textsc{Madgraph5}$. These widths are the source
of the dependence of $n_{\rm ISS}$ on $y_{\nu_{e}}$. In particular, they
scale as $y_{\nu_{e}}^{2}$ when $m_{h}>m_{N}$, and as $y_{\nu_{e}}^{4}$
when $m_{h}<m_{N}$, i.e., the heavy neutrino that mediates the decays
is off-shell. Therefore, as $N$ becomes heavier than the Higgs
boson we expect the bound to quickly become less stringent.
For completeness, we report our computation of the width $\Gamma_{\rm ISS}$  and of the partial widths 
$\Gamma(h\to\bar{\nu}N_{e}+{\rm c.c.}\to\bar{\nu}_{e}\ell_{j}\bar{\ell}_{k}\nu)$ in Appendix A.

To derive the bound we compute the maximal $y_{\nu_{e}}$ such that
\[
n_{\rm ISS}(m_{N},y_{\nu_{e}})<n_{95}(A)
\]
where $n_{95}(A)$ is the 95\% CL limit on the number of events after
the selection from the analysis under consideration. From~\cite{CMS-Collaboration:2012vn}
we extracted $n_{95}(A_{120})=55.9$ and $n_{95}(A_{130})=78.6\,$.
We take as the final bound, which we denote as $y_{\nu_{e},\,95}$, the most stringent one between the
two bounds obtained for the analysis $A_{120}$ and $A_{130}$. It turns out that the bounds derived using 
the analysis $A_{120}$ are stronger for all the cases we have considered. 

\begin{figure}
\begin{centering}
\includegraphics[width=0.6\linewidth]{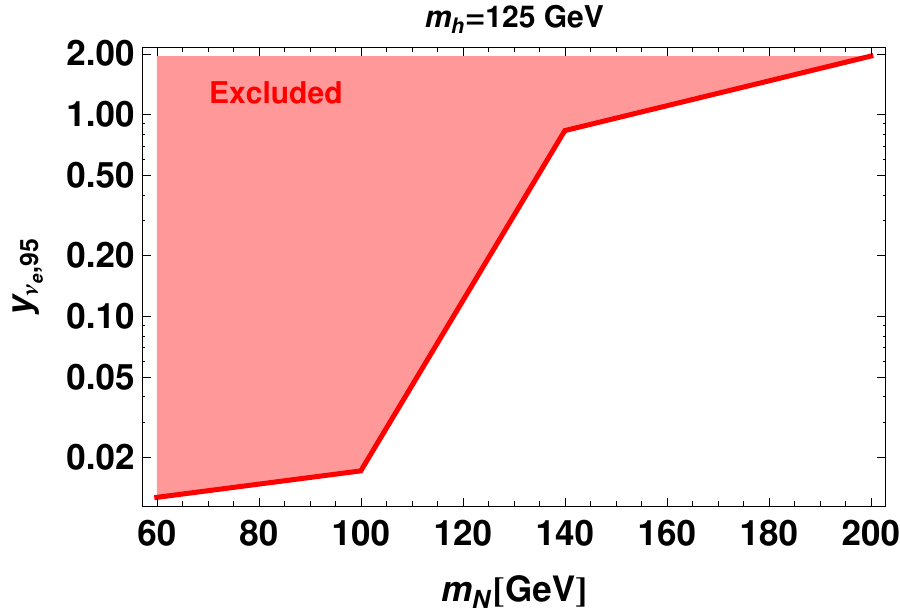}
\par\end{centering}

\caption{\label{fig:Bound125}Bound on $y_{\nu_{e}}$ for $m_{h}=125$
GeV as a function of the mass of the heavy neutrino $N_{e}$.}

\end{figure}

The obtained bound for a fixed $m_{h}=125$ GeV as a function of $m_{N}$
is shown in Figure \ref{fig:Bound125}. 
For the cases where $m_{h}>m_{N}$, we exclude $y_{\nu_{e}}\gsim 0.01$ while for $m_{h}<m_{N}$ 
couplings $y_{\nu_{e}}\gsim 1$ are excluded.

%

\section{Bounds from the observation of a Higgs-like particle at the LHC \label{postdiscovery}}
Evidence of a new particle has been observed in the 2011 and 2012 LHC data \cite{HiggsJuly2012,CMS-PAS-HIG-12-020,ATLAS-CONF-2012-093}. The
region of phase space where the excesses are concentrated suggests
that they are originated by a Higgs-like scalar particle with mass 125 GeV.
Assuming that the new observed state is indeed the Higgs boson which is also involved in the seesaw, further bounds can be obtained from
a global study of the properties of the new particle instead of  just
using the bound on the rate of new phenomena in the $\ell\bar{\ell}\nu\bar{\nu}$
channel. 

In what follows, we use the  measured properties of the new particle  and we shall illustrate how to use the ISS prediction
to put a bound on the size of $y_{\nu_{e}}$. 

The presence of additional decay modes $h\to\bar{\nu} N + \bar{N}\nu$
changes the properties of $h$ in several respects. In fact the total
width of $h$ is increased w.r.t to the SM value. Deviations from
the SM value of the total width are potentially observable in a line-shape
analysis (when $y_{\nu_{e}}$ is large enough) or in a global analysis
of Higgs decay data \cite{Barger:2012qe}.

The change of the total width in turn affects all the branching fractions
of $h$. In particular the rates of modes that do not get contributions
from the new decay mode (such as $\gamma\gamma$, $f\bar{f}$ and $4\ell$)
are suppressed by a factor 
\[
\gamma_{\rm ISS}=\frac{\Gamma_{\rm SM}}{\Gamma_{\rm SM}+\Gamma_{\rm ISS}}\,,
\]
where $\Gamma_{\rm ISS}$ is given in Eq.~(\ref{GammaISS}).

The Higgs decay mode $h\to\ell\bar{\ell}\nu\bar{\nu}$ instead  gets
a contribution from the decays in Eq.~(\ref{eq:ISSdecays}) and its rate
is enhanced by a factor 
\begin{equation}
\mu_{\llnunu}(y_{\nu_{e}})=\frac{n_{\rm ISS}(m_{N},y_{\nu_{e}})}{n_{\rm ISS}(m_{N},0)}\,,  \label{muISS}
\end{equation}
where $n_{\rm ISS}$ is given in Eq.~(\ref{eq:nISS}). 


Altogether the ISS model, compared to the SM, predicts a suppression
by a factor $\gamma_{\rm ISS}$ in the observed rates of all channels
but $\ell\bar{\ell}\nu\bar{\nu}$ %
\footnote{Here we are disregarding the ISS enhancement of final states $jj \nu \bar{\nu} $ and $\ell \bar{\nu}jj $ which, for a light Higgs boson, are less sensitive due to large backgrounds.} which instead is enhanced by the new decays.

The constraint that comes from the increase of the total width is far-reaching. In fact, it also applies to seesaw models where, for any reason, the mode $\llnunu$ does not get enhanced. This might happen, for instance, when the $N$ is very light, say $m_{N}\lesssim 20 \gev$, such that the leptons from the $N$ decay do not pass the selection cuts in Table \ref{tab:Baseline-selection}.  Furthermore, the bound coming from the extra contribution to the width applies to other models  as the case of ISS models where the coupling $y_{\nu_{\tau}}$ dominates the decay. In this case the searches into $\llnunu$ are much less effective, still the effect on the total width provides a bound on the seesaw coupling.

The constraints from the $\llnunu$ channel are more specific to each model. Also they depend on the details of each of the $WW$ analysis. In fact to make a correct use of the measured best-fit signal strengths of the $WW$ channels one should consider each $WW$ analysis separately and compute in detail the efficiencies as we have done for the analysis of~\cite{CMS-Collaboration:2012vn} discussed in Section \ref{llnunubound} .

For these reasons we shall consider the bound from two sets of measurements. In both cases we take the ISS model as reference. In the first case we put a bound on $y_{\nu_{e}}$ coming from all the measured signal strengths excluding the $WW$ channels. This is the safest possibility as we are making the least number of assumptions on the structure of the seesaw couplings. Additionally  we compute the bound using all the available data on the Higgs-like particle at 125 GeV. 
To deal with the several $WW$ analyses  we  make the simplifying assumption that all the efficiencies for the  processes mediated by the seesaw coupling are the same as those computed for the analysis of Section \ref{llnunubound} \footnote{While this is not completely rigorous we expect it to be a good approximation. In fact, the new CMS analysis \cite{CMS-PAS-HIG-12-017} is very similar to~\cite{CMS-Collaboration:2012vn}. Furthermore, one should note that the bound on the seesaw coupling is mostly sensitive to the ratio $\epsilon_{\rm SM}/\epsilon_{jk}$ in Eq.~(\ref{eq:nISS}). As we find in section \ref{llnunubound} 
that the ISS efficiencies are quite similar to those of the SM, 
one can expect our simplifying assumption to be reliable for the ATLAS 
analysis as well.}.


To put a bound we proceed as follows. For each searched decay mode
of the Higgs the collider experiments give the best fit value of $\left(\sum_{p}\sigma_{p}\right)\times BR_{d}$
where $p$ runs on the Higgs production modes and $BR_{d}$'s are the
branching fraction for the various Higgs boson decay modes $d$. These measured
best fits are expressed in units of the SM prediction and are referred
to as best-fit signal strengths
\[
\mu_{d}^{*}\pm\delta\mu_{d}^{*}\,,
\]
where by $\mu_{d}^{*}$ we mean the central value and by $\delta\mu_{d}^{*}$ we mean the symmetrized 1$\sigma$ error on the best-fit of the channel $d$ given by the experiments.

To place a bound we confront the ISS prediction with the latest best-fit
signal strengths given by the experiments at the LHC \cite{HiggsJuly2012,CMS-PAS-HIG-12-020,ATLAS-CONF-2012-093,ATLAS-CONF-2012-091,ATLAS-CONF-2012-092,ATLAS-Collaboration:2012fv,CMS-PAS-HIG-12-015,CMS-PAS-HIG-12-008} and the TeVatron~\cite{The-TEVNPH-Working-Group:2012uq}.
For convenience of the reader the best-fit signal strengths used in
our analysis are reported in Table~\ref{strengths}.
\begin{table}
\begin{tabular}{||c|c||}\hline\hline
\text{CMS $\gamma \gamma $ 2011+2012} & $1.6\pm 0.4$ \\
 \text{CMS ZZ 2011+2012} & $0.8\pm 0.4$ \\
 \text{CMS WW 2011+2012} & $0.6\pm 0.4$ \\
 \text{ATLAS $\gamma \gamma $ 2011+2012} & $1.4\pm 0.5$ \\
 \text{ATLAS ZZ 2011+2012} & $1.3\pm 0.6$ \\
 \text{ATLAS WW 2011} & $0.6\pm 0.6$ \\
 \text{CMS bb AP 2011+2012} & $0.1\pm 0.6$ \\
 \text{CMS $\tau \tau $ 2011+2012} & $-0.2\pm 0.8$ \\
 \text{ATLAS bb AP 2011} & $0.5\pm 2.0$ \\
 \text{ATLAS $\tau \tau $ 2011} & $0.2\pm 1.8$ \\
 \text{TeVatron bb AP} & $2.1\pm 0.7$ \\
 \text{TeVatron WW} & $0.0\pm 1.0$ \\
 \text{CMS WW AP} & $-1.7\pm 3.5$ \\
 \text{CMS $\gamma \gamma $ Dijet 2011} & $4.2\pm 2.0$ \\
 \text{CMS $\gamma \gamma $ Dijet Tight 2012} & $1.3\pm 1.6$ \\
 \text{CMS $\gamma \gamma $ Dijet Loose 2012} & $-0.6\pm 2.0$ \\
\hline\hline
\end{tabular}
\caption{  \label{strengths} Signal strength best-fits extracted
from~\cite{HiggsJuly2012,CMS-PAS-HIG-12-020,ATLAS-CONF-2012-093,ATLAS-CONF-2012-091,ATLAS-CONF-2012-092,ATLAS-Collaboration:2012fv,The-TEVNPH-Working-Group:2012uq,CMS-PAS-HIG-12-015,CMS-PAS-HIG-12-008}. AP stands for associated production. The
best-fit for the AP and Dijet analyses are extracted from \cite{CMS-PAS-HIG-12-015,CMS-PAS-HIG-12-008} . We consistently take the best-fit signal strength for $m_{h}=125$
GeV throughout.} 
\end{table}

As in the ISS model the couplings of the Higgs to the fermions, gluons
and to the gauge bosons are not modified the production cross-sections
are the same as in the SM. This allows us to compute the changes in
the rates from  the changes in the branching fractions only.

From the ISS predictions for the signal strengths $\mu_{d}(y_{\nu_{e}})$
and the measured signal strengths we compute the $\chi^{2}$ for several
choices of $m_{N}$ as a function of the ISS coupling $y_{\nu_{e}}$:

\[
\chi^{2}=\sum_{d}\frac{\left(\mu_{d}(y_{\nu_{e}})-\mu_{d}^{*}\right)^{2}}{\left(\delta\mu_{d}^{*}\right)^{2}}\,,
\]
where $d$ runs on all the measured rates under consideration, and for the ISS $\mu_{\llnunu}$ is given by Eq.~(\ref{muISS}) and $\mu_{d}=\gamma_{ISS}$ for all the other channels. 

\begin{figure}
\begin{centering}
\includegraphics[width=0.65\linewidth]{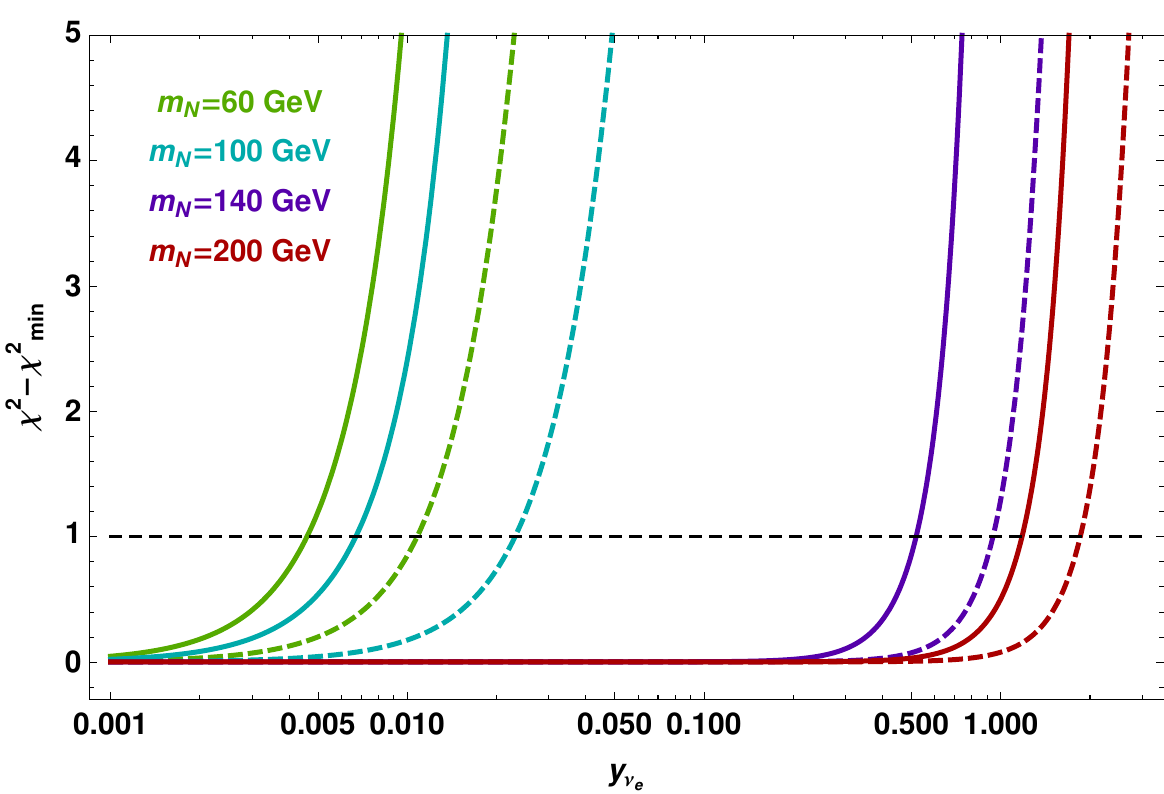}
\par\end{centering}
\caption{\label{chisquare} The $\chi^{2}$ as function of $y_{\nu_{e}}$ in
the ISS model for $m_{h}=125$ GeV and $m_{N}=60,\,100,\,140\,$ and
$200$ GeV. The horizontal black dashed line corresponds to $\chi^{2}=\chi_{\rm min}^{2}+1\,$. The colored  solid lines are for the $\chi^{2}$ of the entire dataset, and the dashed lines are for the $\chi^{2}$ without the $WW$ channels.}
\end{figure}

For each $m_{N}$ the $\chi^{2}$ is minimized w.r.t. $y_{\nu_{e}}$
at a value $\chi_{\rm min}^{2}$. Considering all channels but those with $h\to WW$ we get $\chi^{2}_{\rm min} \simeq 16$, adding the $WW$ channels we get $\chi^{2}_{\rm min}\simeq 20$. 
In all cases we find that the $\chi^{2}$
is minimal for $y_{\nu_{e}}=0$, therefore we derive a 67\% CL upper-bound
on $y_{\nu_{e}}$.  In Figure~\ref{chisquare} we show the obtained
$\chi^{2}-\chi^{2}_{\rm min}$ with and without the $WW$ channels.
 The inclusion of the $WW$ channels improves the bound by a factor of about 2 in most cases. 

\section{Discussions and conclusion \label{conclusion}}
In this paper we have derived  bounds on the Dirac Yukawa couplings of the lepton doublet in inverse seesaw models for neutrino masses using LHC Higgs data. In  generic versions of these models, such bounds are useful since  one could understand small neutrino masses while keeping the Yukawa couplings to be of order one and the singlet fermion masses in the 100 GeV range. We have focused on the cases where the electron or muon Yukawas are the dominant ones and also discussed the case with flavor-democratic Yukawa. Our discussion applies to the supersymmetric version of the model as well. It is perhaps worth pointing out that in SUSY ISS model, there are additional $D$-term contributions~\cite{an} of order of a few GeV as well as new F-term contributions~\cite{el}  to the Higgs mass thus relieving some MSSM parameter space. We find that for singlet fermion masses between $60-140$ GeV, useful bounds can be derived on the Yukawa couplings from the recent LHC data on Higgs searches.  

It is also worth noting that in the low-scale type-I and inverse seesaw models, there are limits on the mixing parameter $\frac{yv}{M}$ from leptonic unitarity~\cite{LU} and lepton flavor violation~\cite{lfv}. The current bounds for the electron-flavor is 
$\frac{yv}{\sqrt{2}M}\lsim 0.044$ and for the muon sector it is $\lsim 0.03$ (see \cite{LU,lfv} for details). These bounds are weaker than what we obtain in this paper for $M\sim 100$ GeV 
from LHC data.

It is also worth pointing out that if we assumed a pattern for Dirac Yukawa couplings similar to the charged fermion case i.e. $y_{\nu_\tau} \gg y_{\nu_\mu}, y_{\nu_e}$, then the dominant mode for $h$ decay will involve the $\tau$ decay and our constraints will not apply in a straightforward manner. However, the constraints from the global fit due to the increase of the total width shown as dashed lines in Figure~\ref{chisquare} will still apply to $y_{\tau}$.

\section*{Acknowledgment}

RF thanks Alessandro Strumia for discussions on the LHC Higgs data. 
The work of BD and RNM is supported in part by the National Science Foundation Grant Number PHY-0968854.  The work of RF is supported  by the NSF under grants PHY-0910467 and PHY-0652363, and by the Maryland Center for Fundamental Physics. This work of BD was partly supported by the Lancaster-Manchester-Sheffield Consortium for Fundamental Physics under STFC grant ST/J000418/1.  BD acknowledges the local 
hospitality and computing facilities provided at IACS, Kolkata, during the 
final stages of this work. 

\section*{Note added in proof}
After our paper was submitted, the 8-TeV LHC data on the final state $\ell \nu \bar\ell\bar\nu$ appeared~\cite{CMS-PAS-HIG-12-017}. 
Compared to the 7-TeV analysis, the major differences in the 8-TeV analysis are:
\begin{enumerate}
\item $p_T^{\ell,{\rm min}}>10$ GeV for both same- and opposite-flavor leptons. 
\item $m_{\ell\ell}>12$ GeV for both same- and opposite-flavor leptons. 
\item $E_T^{\rm miss}>20$ GeV for both same- and opposite-flavor leptons.  
\end{enumerate}
Using these new cuts, we repeated the derivation of the bounds along the lines of the method described in the text. For a SM Higgs of 125 GeV we find slightly more stringent bounds, that improve on the 7-TeV results, by roughly 10-20\%.

After this paper was posted on the arXiv, another paper~\cite{okada} 
studying the collider 
signatures of ${\cal O}(100)$ GeV pseudo-Dirac neutrinos in the inverse seesaw scenario was posted. 
We thank the referee for bringing this paper to our attention.
\appendix
\section{Decay Widths of the Heavy Neutrino}

In this Appendix, we collect the partial and total widths of the heavy neutrino $N$ (Table~\ref{tab:mnlmh1}) as well as the partial widths of the Higgs and the increment in its total width (Tables~\ref{tab:mnlmh2} , \ref{tab:mngmh2} and \ref{tab:mngmh1}) that have been used to compute the event yield and the bounds in Sections  
\ref{llnunubound} and \ref{postdiscovery}. 

\begin{table}[h!]
	\begin{center}
		\begin{tabular}{||c||c|c|c|c||}\hline\hline
$m_N$ & $\Gamma(N_e\to e^-e^+\nu_e)$ & $\Gamma(N_e\to \nu_e \mu^-\mu^+)$ & $\Gamma(N_e\to e^-\mu^+\nu_\mu)$ & $\Gamma(N_e)$ \\ 
(GeV) & $[y_N^2\cdot$(GeV)] & $[y_N^2\cdot$(GeV)] & $[y_N^2\cdot$(GeV)] & $[y_N^2\cdot$(GeV)] \\ \hline\hline
60 & $1.464\times 10^{-4}$& $2.569\times 10^{-5}$ & $2.32\times 10^{-4}$ & 0.002716\\ \hline 
100 &0.03204 & 0.001495 & 0.03182 & 0.3263 \\ \hline \hline
		\end{tabular}
	\end{center}
	\caption{The relevant partial widths and the total width of $N_e$ for $m_N<m_h$ in the ISS model with dominant $y_{\nu_{e}}$. 
	}
	\label{tab:mnlmh1}
\end{table}
\begin{table}[h!]
	\begin{center}
		\begin{tabular}{||c|c||c|c|c|c||}\hline\hline
			$m_h$ & $m_N$ & $\Gamma_{\rm ISS}$ & $\Gamma_{\rm ISS}(e^+e^-)$ &  $\Gamma_{\rm ISS}(\mu^+\mu^-)$ & $\Gamma_{\rm ISS}(e^\mp\mu^\pm)$  \\ 
			(GeV) & (GeV) & [$y_N^2\cdot$(GeV)] & [$y_N^2\cdot$(GeV)] & [$y_N^2\cdot$(GeV)] & [$y_N^2\cdot$(GeV)]\\					\hline\hline
			125 & 60 & 2.9458 & 0.1588 &0.0279 & 0.2516 \\ \cline{2-6}
			& 100 & 0.6446 & 0.0633 & 0.0030 & 0.0629\\ \hline 
			
			\hline\hline
		\end{tabular}
	\end{center} 
	\caption{The relevant  widths  $h\to \llnunu$ and the increment of the total decay width of the Higgs in the ISS model with dominant $y_{\nu_{e}}$ and $m_N<m_h$. The partial widths are calculated by multiplying the total width $\Gamma_{\rm ISS}$  
	by the branching fractions of $N$ computed from Table~\ref{tab:mnlmh1}. }
	\label{tab:mnlmh2}
\end{table}

		\begin{table}[h!]
	\begin{center}
		\begin{tabular}{||c|c||c|c|c||}\hline\hline
			$M_h$ & $M_N$ & $\Gamma(h\to \bar\nu_e W^+e^-)$ 
			& $\Gamma(h\to \bar\nu_e Z\nu_e)$  & $\Gamma_{\rm ISS}(h)$ \\ 
				(GeV) & 
			(GeV) & 
			$[y_N^4\cdot({\rm GeV})]$ & 
			$[y_N^4\cdot({\rm GeV})]$ & 
			$[y_N^4\cdot({\rm GeV})]$ \\ 
			\hline\hline
			125 & 140 & $1.658\times 10^{-4}$ & $1.051\times 10^{-4}$ & $5.42\times 10^{-4}$ \\ \cline{2-5}
			& 200 & $1.119\times 10^{-5}$ & $6.655\times 10^{-6}$ & $3.57\times 10^{-5}$\\ \hline
\hline
		\end{tabular}
	\end{center}
	\caption{Total increment of the decay width of the Higgs in the ISS model with dominant $y_{\nu_{e}}$ for $m_N>m_h$. }
	\label{tab:mngmh2}
\end{table}

		\begin{table}[h!]
			\begin{center}
				\begin{tabular}{||c|c||c|c|c||}\hline\hline
					$m_h$ & $m_N$ & $\Gamma_{\rm ISS}(e^+e^-)$ & $\Gamma_{\rm ISS}(\mu^+\mu^-)$ & $\Gamma_{\rm ISS}(e^\mp\mu^\pm)$ \\
					(GeV) & (GeV) & 
					$[y_N^4\cdot({\rm GeV})]$ & 
					$[y_N^4\cdot({\rm GeV})]$ & 
					$[y_N^4\cdot({\rm GeV})]$ \\ \hline\hline
					125 & 140 & $ 3.882\times 10^{-5} $& 
					$3.681\times 10^{-6} $ & $ 3.53\times 10^{-5} $\\ \cline{2-5}
					& 200 & $2.657\times 10^{-6} $ & $ 2.433\times 10^{-7}$ & $2.416\times 10^{-6} $\\ \hline \hline
				\end{tabular}
			\end{center}
			\caption{The relevant decay widths $h\to \ell \bar{\ell} \nu\nu$ for $m_N>m_h$ in the ISS model with dominant $y_{\nu_{e}}$.}
			\label{tab:mngmh1}
		\end{table}



\end{document}